\documentstyle[preprint,aps]{revtex}
\input psfig.sty
\tightenlines
\begin{document}
\bibliographystyle{Unsrt}
\title{Charged-particle pseudorapidity density distributions
from Au+Au collisions at $\sqrt{s_{NN}}$=130 GeV}


\author{ B.B.Back$^1$, M.D.Baker$^2$, 
D.S.Barton$^2$, R.R.Betts$^{3}$, R.Bindel$^4$,  
A.Budzanowski$^5$, W.Busza$^{6}$, A.Carroll$^2$,
M.P.Decowski$^6$, 
E.Garcia$^4$, N.George$^1$, K.Gulbrandsen$^6$, 
S.Gushue$^2$, C.Halliwell$^3$, 
G.A.Heintzelman$^2$, C.Henderson$^6$, D.J.Hofman$^3$,
R.Ho\l y\'{n}ski$^5$, B.Holzman$^{2,3}$, 
E.Johnson$^7$, J.L.Kane$^6$, J.Katzy$^6$, N. Khan$^7$, W.Kucewicz$^3$, 
P.Kulinich$^6$,
W.T.Lin$^8$, S.Manly$^7$,  D.McLeod$^3$, J.Micha\l owski$^5$,
A.C.Mignerey$^4$, J.M\"ulmenst\"adt$^{6}$, R.Nouicer$^3$, 
A.Olszewski$^{2,5}$, R.Pak$^2$, I.C.Park$^7$, 
H.Pernegger$^6$, C.Reed$^6$, L.P.Remsberg$^2$, 
M.Reuter$^3$, C.Roland$^6$, G.Roland$^6$, L.Rosenberg$^6$, 
P.Sarin$^6$, P.Sawicki$^5$, 
W.Skulski$^7$, 
S.G.Steadman$^6$, 
P.Steinberg$^2$, G.S.F.Stephans$^6$,  M.Stodulski$^5$, A.Sukhanov$^2$, 
J.-L.Tang$^8$, R.Teng$^7$, A.Trzupek$^5$, 
C.Vale$^6$, G.J.van Nieuwenhuizen$^6$, 
R.Verdier$^6$, B.Wadsworth$^{6}$, F.L.H.Wolfs$^7$, B.Wosiek$^5$, 
K.Wo\'{z}niak$^5$, 
A.H.Wuosmaa$^1$, B.Wys\l ouch$^6$\\
(PHOBOS collaboration)}

\address{
$^1$ Physics Division, Argonne National Laboratory, Argonne, IL 60439-4843\\
$^2$ Chemistry and C-A Departments, Brookhaven National Laboratory, Upton, NY 11973-5000\\
$^3$ Department of Physics, University of Illinois at Chicago, Chicago, IL 60607-7059\\
$^4$ Department of Chemistry and Biochemistry, University of Maryland, College Park, MD 20742\\
$^5$ Institute of Nuclear Physics, Krak\'{o}w, Poland\\
$^6$ Laboratory for Nuclear Science, Massachusetts Institute of Technology, Cambridge, MA 02139-4307\\
$^7$ Department of Physics and Astronomy, University of Rochester, Rochester, NY 14627\\
$^8$ Department of Physics, National Central University, Chung-Li, Taiwan
}

\date{\today}          
\maketitle                     

\begin{abstract}

The charged-particle pseudorapidity density $dN_{ch}/d\eta$ has been
measured for Au+Au collisions at $\sqrt{s_{NN}}$=130 GeV at RHIC, using the
PHOBOS apparatus.  The total number of charged particles produced for the
3\% most central Au+Au collisions for $|\eta|\leq$5.4 is found to be
4200$\pm$470.  The evolution of $dN_{ch}/d\eta$ with centrality is
discussed, and compared to model calculations and to data from
proton-induced collisions. The data show an enhancement in charged-particle
production at mid-rapidity, while in the fragmentation regions, the results
are consistent with expectations from $pp$ and $pA$ scattering.
\end{abstract}


Nucleus-nucleus collisions at highly relativistic energies from the
Relativistic Heavy Ion Collider (RHIC) present a new opportunity for the
study of strongly interacting matter under conditions of very high
temperature and energy density. Already, data from central Au+Au collisions
at energies of $\sqrt{s_{NN}}$=56 and 130 GeV \cite{PRL} have shown an
increase in mid-rapidity particle production over that observed in Pb+Pb
collisions at the CERN SPS \cite{SPS}. In addition, the charged-particle
yields, when scaled by the number of participating nucleons ($N_{part}$),
exceed the values observed in proton-proton collisions at similar
$\sqrt{s_{NN}}$. These results have been used to infer an energy density at
the highest value of $\sqrt{s_{NN}}$ that is considerably larger than was
achieved at CERN, and which is well within the region where new physics is
to be expected.

The distribution of $dN_{ch}/d\eta$ over the entire range of pseudorapidity
$\eta$, where $\eta=-ln(\tan(\theta/2))$ and $\theta$ is the angle with
respect to the beam, represents a time-integral of particle production
throughout the collision, and reflects all effects that contribute to the
production of charged particles. In addition to the influence of both hard
and soft production processes, $dN_{ch}/d\eta$ is sensitive to nuclear
effects in the initial parton distributions, as well as to effects from
hadronic re-interactions in the final state.

The data for $dN_{ch}/d\eta$ at mid-rapidity ($|\eta|<1$)
for the most
central Au+Au collisions at $\sqrt{s_{NN}}$=56 and 130 GeV\cite{PRL} are in
reasonable agreement with the predictions of a number of models including
HIJING\cite{HIJING}, a saturation model(EKRT)\cite{EKRT}, and purely
hadronic models (e.g. LUCIFER\cite{LUCIFER,LUCIFER2}). The centrality
dependence of $dN_{ch}/d\eta$ at mid-rapidity has also been
measured\cite{PHENIX,peter}. These results suggest some deviation from both
the HIJING and EKRT model predictions, although they are in broad agreement
with the results of calculations by Kharzeev and Nardi\cite{KN}.

An extension of the measurements of $dN_{ch}/d\eta$ data beyond
mid-rapidity, for a range of impact parameters, is necessary to further
constrain models.  It is also of interest to determine whether the observed
scaling of the charged-particle yield with $N_{part}$ at mid-rapidity
is modified at large values of $\eta$ where, in proton-nucleus ($pA$)
collisions\cite{busza,demarzo,Barton}, re-scattering, stopping, and target
fragmentation influence the shape of the $dN_{ch}/d\eta$ distributions.  In
order to address these questions, we have used the PHOBOS apparatus to
measure the charged-particle pseudorapidity density $dN_{ch}/d\eta$ from
Au+Au collisions at $\sqrt{s_{NN}}$=130 GeV over the range $|\eta|<5.4$.

The PHOBOS experiment at RHIC largely consists of several arrays of
silicon pad detectors.  The details of the experimental arrangement are
described in Ref.~\cite{BBB}.  The procedures used for event selection, the
determination of the collision-vertex position, event centrality, and the
estimation of $N_{part}$, have been described in Refs.~\cite{PRL,peter}. The
specific elements of the experiment used in the current measurement and the
analysis procedures leading to $dN_{ch}/d\eta$ are described below.

The data samples included in the current analysis were taken at a collision
energy of $\sqrt{s_{NN}}$=130 GeV.  The collision vertices were confined to
a region within $\pm$10 cm of the nominal beam crossing and center of the
apparatus ($z=0$). At mid-rapidity, charged particles were detected, and
their energy deposition measured, with an octagonal array of pad detectors
approximately 1 m long (the ``Octagon'') that surrounds the thin-walled Be
beam pipe. The Octagon subtends the full azimuthal range, except for regions
where sensors that would intercept the acceptance for the
tracking spectrometers and vertex finding detectors are removed. For
collision vertices within $|z|<$10 cm, the pseudorapidity 
coverage of the Octagon is
complete for $|\eta |\leq$3.2.  Six rings of silicon pad detectors placed at
distances of $|z|\approx$1, 2 and 5 m, detected particles in pseudorapidity
ranges of $3\leq|\eta|\leq 4$, $4\leq|\eta|\leq 4.7$, and $4.7\leq|\eta|\leq
5.4$, respectively. Thus, for collisions within the 20 cm long region in the
center of the experiment, there are no significant gaps in the $\eta$
coverage from one sub-detector to the next. The total numbers of pads in the
Octagon and Rings are 11,040 and 3,072, respectively.

Two complementary methods were used to analyze the pseudorapidity
distribution data. The first method (``Hit Counting'') uses the segmentation
of the multiplicity detector.  After merging of signals in neighboring pads,
in cases where a particle travels through more than a single pad, the
deposited energy was corrected for the angle of incidence, so that all
tracks originating from the collision vertex possess a common average value
of the deposited energy ($\Delta E \approx 80$ keV).  Pads containing more
than 75\% of this value were counted as occupied.  This requirement largely
suppresses hits from background and from secondary-particles not originating
from the primary collision vertex. Then, for a given value of $\eta$ and bin
$i$ in collision centrality, the number of hit pads $N(\eta,i)$ was
corrected for the effects of multiple occupancy, where more than one
particle travels through a given pad, as well as for contributions from the
remaining secondary particles, absorption in the beam pipe, and weak decays
of primary particles.  The number of particles per pad was determined as a
function of $\eta$ and centrality bin in two ways. In the first method, the
probability $P(N)$ of $N$ particles passing through a given pad was assumed
to be Poisson distributed. The mean occupancy $\mu (\eta,i)$ could then be
determined from the ratio of occupied to unoccupied pads in a range of
$\eta$ for each centrality bin. Second, as a check, the occupancy was also
determined from the energy-deposition spectra.  A fitting procedure was used
to determine the relative contribution to the total energy deposition of one
or more particles, thus providing an independent measure of the mean
occupancy $\mu (\eta ,i)$. Maximum values of approximately 1.6 particles per
hit pad were obtained for the most central collisions at mid-rapidity.

To account for effects or biases not treated in the above analysis
procedures, a final correction was deduced from GEANT simulations of the
detector response using events from the HIJING\cite{HIJING}, RQMD\cite{RQMD}
and VENUS\cite{VENUS} event generators. The ratio between the simulated,
occupancy-corrected $dN_{ch}/d\eta$ distributions and the known ``Truth''
distributions formed this last set of corrections, which ranged up to 15\%
in the Octagon and up to 50\% in the Rings. These background correction
factors, dependent upon both $\eta$ and centrality bin, were applied to the
occupancy-corrected data, yielding $dN_{ch}/d\eta$. The final results
obtained using both hit counting methods were in good agreement.

In the second (``Analog'') method, the pseudorapidity distribution was
extracted directly from the energy deposition $\Delta E(\eta)$ in the
multiplicity detectors. The measured energy deposition was transformed into
$dN_{ch}/d\eta$ using quantities derived from the results of Monte-Carlo
simulations. The average energy per track ($\Delta E_{tr}$) was determined
as a function of $\eta$ using particles from HIJING events, passed through
the GEANT simulation of the detector.  The fraction of primary particles
$f_{prim}(\eta)$ was determined from the same simulations.  Then,
$dN_{ch}/d\eta=\frac{\Delta E(\eta)\times f_{prim}(\eta)}{\Delta
E_{tr}(\eta)\Delta \eta}$.  
Although the two methods differ qualitatively in the way in
which the energy-deposition information from the multiplicity detectors is
used, and rely differently upon the results of Monte-Carlo simulations, they
yield results that differ generally by $\lesssim$5\% throughout the range in
$\eta$, well within the systematic uncertainties (see below), which are
approximately 10\%.

The systematic uncertainty in the occupancy correction for the hit-counting
analysis was obtained by comparing the results from the full analysis chain
using Poisson-derived occupancy corrections with those derived from the
measured $\Delta E$ spectra.  The average deviations are less than 3\%,
yielding a partial systematic error of $\approx$ 3\% at mid-rapidity. The
systematic uncertainties from the Monte-Carlo simulations have been
estimated by using different assumptions in the GEANT simulation, as well as
different event generators, including RQMD\cite{RQMD} and
VENUS\cite{VENUS}.  The variations observed in the derived background
corrections are between 4 and 8\%, suggesting a total systematic uncertainty
of approximately 10\%. The systematic errors in the Analog analysis arise
from uncertainties in the Monte Carlo simulations, as well as those in the
absolute energy calibrations of the Silicon-pad detectors. The latter are
approximately 5\%, yielding a total systematic error for the Analog method
of approximately 10\%, similar to that for the hit-counting method.

\begin{figure}
\centerline{\psfig{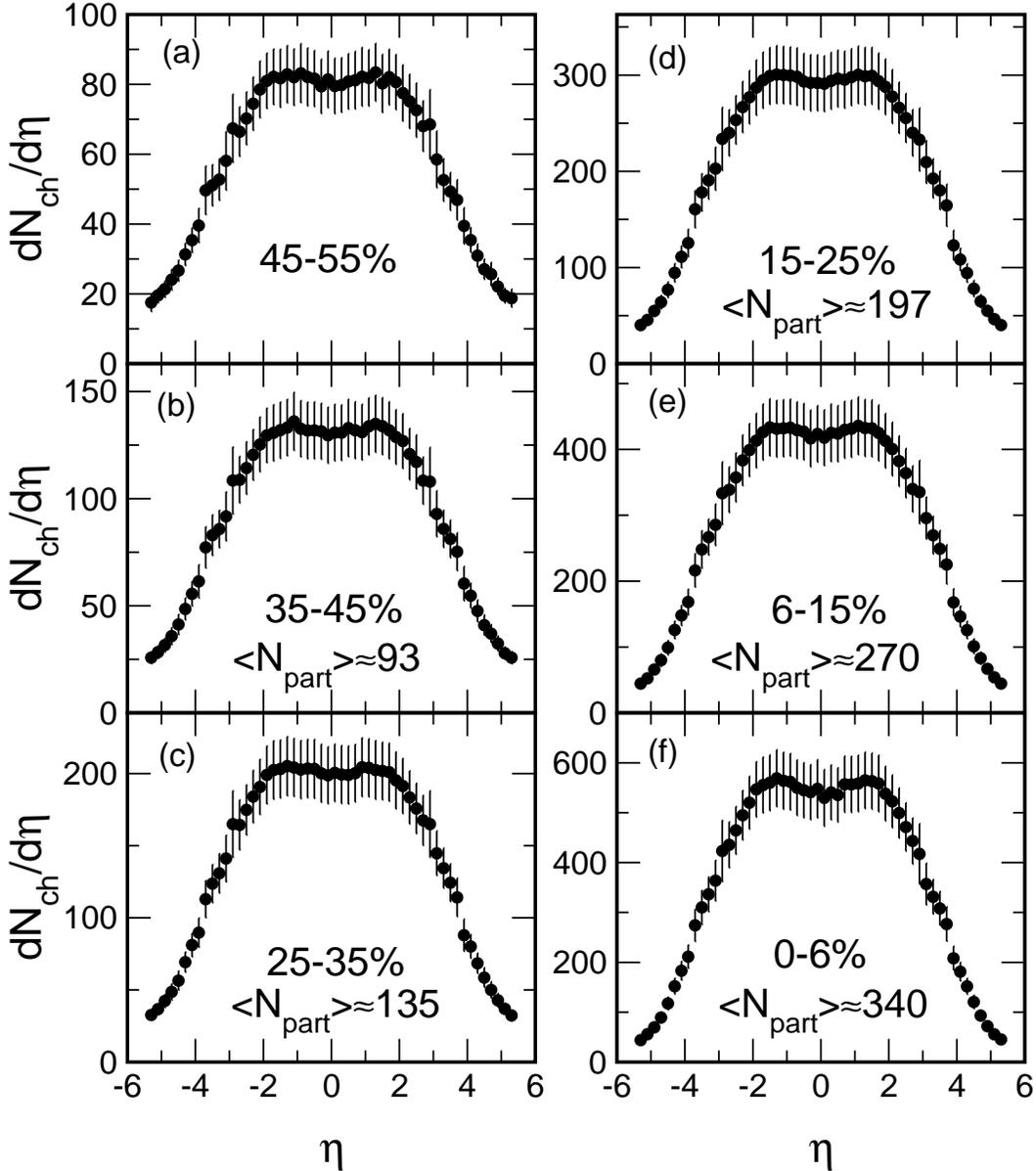}}
\caption{Charged-particle pseudorapidity density
$dN_{ch}/d\eta$ from $\sqrt{s_{NN}}$=130 GeV Au+Au collisions, for different
centrality bins, as defined by different fractions of the total observed
cross section.  The error bars reflect primarily the systematic uncertainties.
The average number of participants $<N_{part}>$ for each bin
is also indicated.  For fractions of the cross section $>45\%$ the systematic
uncertainties in the $<N_{part}>$ determination are still under study and no
value is quoted.}
\label{fig1}
\end{figure}

Our final results are presented in Fig.~\ref{fig1}(a)-(f), which show the
error-weighted average values of $dN_{ch}/d\eta$ from the two procedures for
six different centrality bins.  The error bars represent a convolution of
the estimated systematic errors in the different analyses.  The different
centrality bins are denoted by the corresponding fraction of the observed
total cross section, as well as by the deduced average number of participant
nucleons $<N_{part}>$.  While the occupancy and background corrections
applied to the data from different parts of the apparatus are quite
different in magnitude, the data from the Octagon and Ring detectors merge
smoothly in most cases. The present analyses give values of $dN_{ch}/d\eta$
that are in good agreement from the independent ``Tracklet'' analysis
presented in
\cite{peter}.  For example, for the 6\% most central collisions, we find
$dN_{ch}/d\eta|_{|\eta|\leq1}=547\pm55$, compared to 580$\pm$25 from
Ref.\cite{peter}.

\begin{figure}
\centerline{\psfig{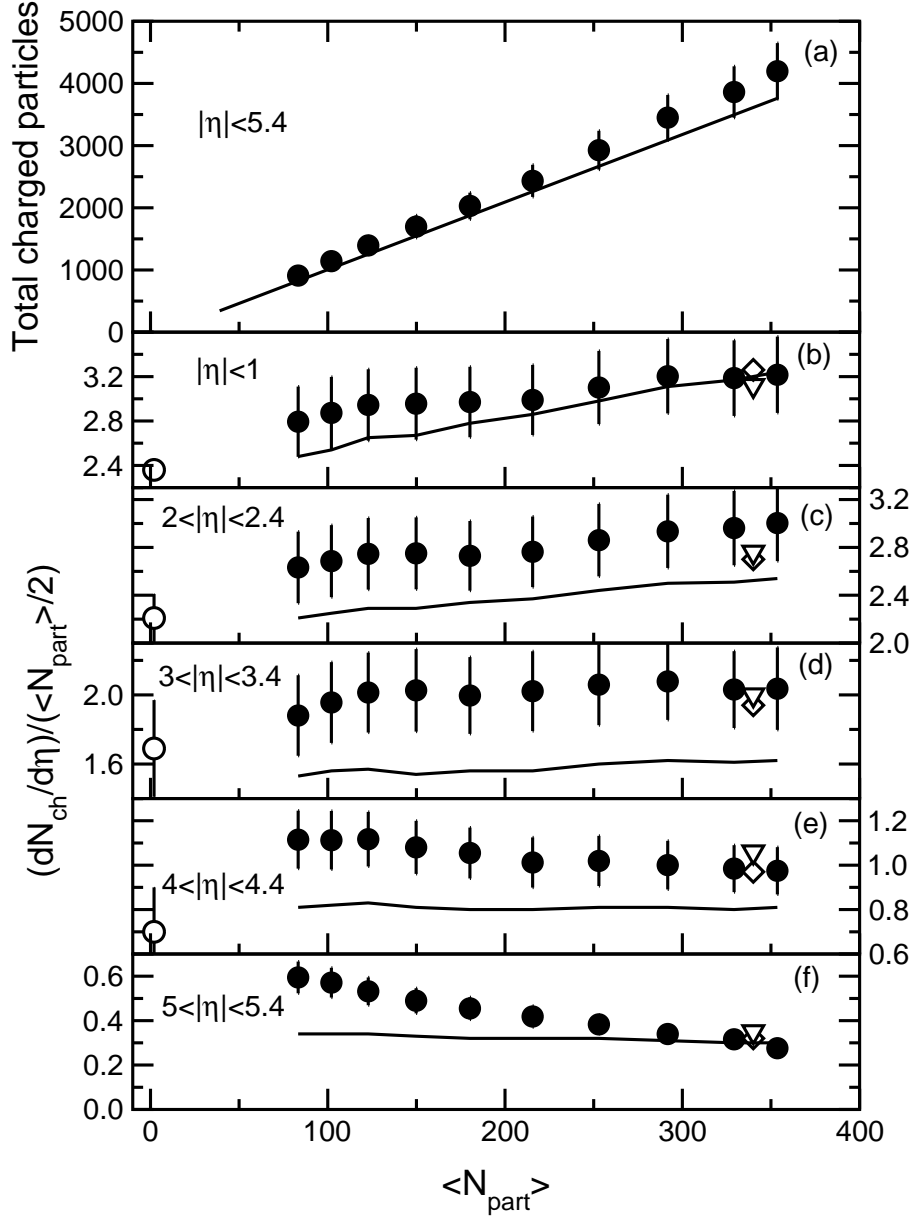}}
\caption{(a): Total number of charged particles detected within
the range $-5.4<\eta <5.4$.  The symbols are the measured data, and the
line represents the prediction of the code HIJING.  The error bars reflect
systematic uncertainties. (b)-(f): Centrality dependence of $dN_{ch}/d\eta$
for different ranges of $|\eta|$.  The filled symbols are the measured data,
and the solid curves are the HIJING predictions.  The open diamonds and
triangles refer to the predictions of the AMPT, and LUCIFER, models,
respectively for the 6\% most central collisions. The open circles represent
the values from $pp$ and $p\overline{p}$ collisions.}
\label{fig2}
\end{figure}

The integral of the distributions $N_{ch}^{tot}$, plotted in
Fig.~\ref{fig2}(a) as a function of centrality, is a direct measure of the
total entropy produced in the collisions.  Predictions of its magnitude have
varied by as much as factor of two \cite{Bass}.  With increasing $\langle
N_{part} \rangle$, the observed values of $N_{ch}^{tot}$ change smoothly
from $910\pm100$ for the 40-45\% centrality bin to $4200\pm470$ for the 3\%
most central collisions.  Per participant pair, these numbers correspond
respectively to 21.8$\pm2.6$ and 23.7$\pm$2.7, compared with the $pp$ and
$p\overline{p}$ non-diffractive total charged particle multiplicity of
18.5$\pm0.7$\cite{wroblewski}. The predictions of the HIJING model reproduce
the general trend of the $N_{ch}^{tot}$ centrality dependence, but
systematically under-predict the observed values by approximately 10\%.

The shapes of the pseudorapidity distributions evolve gradually with
increasing centrality, as shown in Fig.~\ref{fig1}(a-f). For all centrality
bins, there is a plateau region between $-2 < \eta < 2$, followed by a rapid
drop-off towards larger pseudorapidities. A more detailed study of the
centrality dependence of the shape is given in Figs.~2(b-f) and Fig~3(a). In
Figs.~2(b-f), the $N_{part}$ dependence of $dN_{ch}/d\eta$ normalized per
participant pair $<N_{part}/2>$, is plotted for five pseudorapidity bins
ranging from $|\eta|<1$ to $5<|\eta|<5.4$. Also plotted are data from
$pp$\cite{alner} and $p\overline{p}$\cite{ua5} collisions, scaled as
described below, as open circles, and predictions from the HIJING model as
solid lines.  The statistical uncertainties are small and the systematic
uncertainties are comparable to those described above. Figure 3(a) shows the
pseudorapidity distribution for peripheral (35-45\%) and central (0-6\%)
Au+Au collisions, scaled by the respective number of participants.

For all pseudorapidity bins in Fig.~2(b-f), the data evolve smoothly from
the most peripheral to the most central collisions.  As seen in Fig.~2(b,c)
and Fig.~3(a), central collisions yield a 10-15\% higher $dN_{ch}/d\eta$ per
participant in the plateau region, compared to peripheral events. This
difference decreases at larger pseudorapidities and reaches a turnover point
between $|\eta|=3$ and 4, as seen in Fig.~3(a). Beyond that, a higher yield
per participant is found in peripheral collisions.  In Fig.~2(d) and (e)
this turnover can be seen as a change in slope of the $N_{part}$ dependence
of $dN/d\eta$ in the respective $\eta$ bins. In the highest 1.5 units of
pseudorapidity, as is observed in $p+A$ collisions at lower energies
\cite{busza,demarzo,Barton}, the scaled charged-particle density actually
falls with $N_{part}$, being reduced by nearly a factor of two for $|\eta| >
5$. Qualitatively, the changes in the distributions from peripheral, to
central collisions in the fragmentation regions are similar to those
observed in $pA$ collisions
\cite{busza,demarzo,Barton}.

\begin{figure}
\psfig{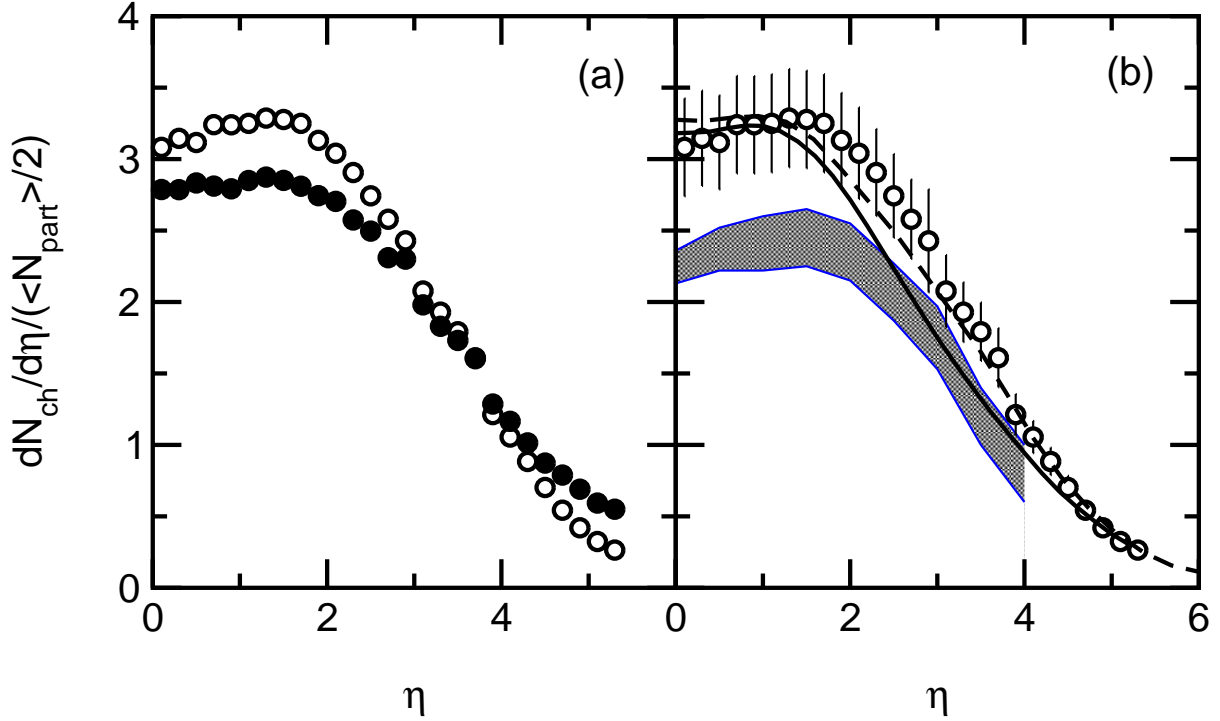}
\caption{(a).
$dN_{ch}/d\eta/(<N_{part}>/2)$ for central (0-6\%, open symbols) and
peripheral (35-45\%, filled symbols) collisions. The error bars have been
suppressed for clarity. (b) Comparison of the measured pseudorapidity
distribution for the 0-6\% centrality bin with the HIJING (thin curve) and
AMPT (dashed curve) models.  The grey band represents data from $pp$ and
$p\overline{p}$ collisions interpolated to $\sqrt{s_{NN}}$=130 GeV.}
\label{fig3}
\end{figure}

Finally, a comparison of the shapes from $pp/p\overline{p}$ collisions and
central Au+Au collisions is given in Fig.~3(b). It shows the scaled
pseudorapidity distribution for central collisions (0-6\%), compared to
scaled data from $pp$ and $p \overline{p}$ collisions, shown as a grey band.
The $pp/p\overline{p}$ distribution was obtained by scaling the measured
$dN/d\eta$ distributions from $pp$ collisions at $\sqrt{s} = 53$~GeV and $p
\overline{p}$ collisions at $\sqrt{s} = 200$ and 546 GeV \cite{alner,ua5}
horizontally by $y_{max}(130~GeV)/y_{max}(\sqrt{s})$, where 
$y_{max}(\sqrt{s}) =
\ln({\sqrt{s}/m_p})$, and vertically using the parameterization of
$dN/d\eta_{|\eta| < 1}$ from
\cite{ua5}\footnote{This empirical scaling is
chosen such that the extrapolation of three sets of $pp$ or $p\overline{p}$
data are confined within the band shown on Fig.~\ref{fig3}(b).}. The grey
band gives an estimate of the uncertainty of this extrapolation procedure.

The scaled $dN_{ch}/d\eta$ is found to be higher in central Au+Au collisions
than in $pp/p\overline{p}$ over the full pseudorapidity range out to $|\eta|
> 4$, with the largest excess observed in the central plateau region. This
is in contrast with the HIJING prediction, which shows an excess in Au+Au
collisions only for $|\eta|\leq 2-3$. A possible origin of this qualitative
difference is suggested by the AMPT model of Zhang et al.\cite{AMPT}. In
this model, the initial state parton distribution is obtained the same way
as in HIJING, but is followed by a parton cascade \cite{Zhang}, string
fragmentation and hadronic re-scattering using a relativistic transport model
\cite{ART}. This model reproduces the excess in particle production at
higher pseudorapidities seen in Au+Au collisions relative to
$pp/p\overline{p}$.  Furthermore, predictions of $dN_{ch}/d\eta$ in a
completely hadronic framework (e.g. LUCIFER)\cite{LUCIFER2} are very similar
to those of the AMPT calculations, and are also in good agreement with the
data for the most central collisions.  These observations suggest that
effects in the hadronic phase should be taken into account to provide a full
description of the data. A comparison of the predicted centrality dependence
of the full distributions from these models with the data would also be of
interest. Further insight will be gained from future RHIC data, which should
include more precise reference data from $pp$ collisions, as well as
nucleus-nucleus data at different collision energies.  These results will
provide a basis for separating the effects of new phenomena from
conventional hadronic physics.

This work was supported in part by US DoE grants
DE-AC02-98CH10886, DE-FG02-93ER-404802, DE-FC02-94ER40818,
DE-FG02-94ER40865, DE-FG02-99ER41099, W-31-109-ENG-38, and NSF grants 9603486,
9722606, and 0072204.  The Polish groups were partially supported by KBN
grant 2P03B 04916. The NCU group was partially supported by the NSC of
Taiwan under contract NSC 89-2112-M-008-024.

\end{document}